\documentclass[11pt]{article}

\usepackage[english]{babel}
\usepackage{amsmath,amssymb}
\usepackage{graphicx}
\usepackage[sort&compress,numbers]{natbib}
\usepackage[colorinlistoftodos]{todonotes}
\usepackage[autostyle]{csquotes}
\usepackage[margin=1in]{geometry}

\newcommand{\ev}[1]{E\left[ #1 \right]}

\newcommand{\evg}[1]{\left\langle #1 \right\rangle}
\newcommand{\evgo}[2]{\left\langle #1 \right\rangle_{#2}}
\newcommand{\est}[1]{\widehat{#1}}
\newcommand{\len}{L}
\newcommand{\Tl}{T_{\len}}
\newcommand{\taul}{\tau_{\len}}

\newcommand{\PTl}{P_{\Tl}}
\newcommand{\pT}{p_T}
\newcommand{\PT}{P_T}
\newcommand{\Pl}{P_{\len}}

\newcommand{\Ptaul}{P_{\taul}}

\newcommand{\ltpT}{\tilde{p}_T}
\newcommand{\ltptaul}{\tilde{p}_{\taul}}
\newcommand{\rblock}{r_\text{block}}

\newcommand{\fig}[1]{Fig.~\ref{fig:#1}}
\newcommand{\figs}[2]{Figs.~\ref{fig:#1} and \ref{fig:#2}}

\title{Minimal-assumption inference from population-genomic data}

\author{Daniel B.~Weissman$^{1,2}$, Oskar Hallatschek$^{2}$\\
$^1$Dept.~of Physics, Emory University, Atlanta, GA 30322\\
$^2$Depts.~of Physics and Integrative Biology, University of California, Berkeley, CA 94720
 }

\begin{document}
\maketitle

\begin{abstract}
Samples of multiple complete genome sequences contain vast amounts of information about the evolutionary history of populations,
much of it in the associations among polymorphisms at different loci. 
Current methods that take advantage of this linkage information rely on models of recombination
and coalescence, limiting the sample sizes and populations that they can analyze.
We introduce a method, Minimal-Assumption Genomic Inference of Coalescence (MAGIC),
that reconstructs key features of the evolutionary history,
including the distribution of coalescence times,
by integrating information across genomic length scales
without using an explicit model of recombination, demography or selection. 
Using simulated data, we show that MAGIC's performance is comparable to PSMC' on single 
diploid samples generated with standard coalescent and recombination models. 
More importantly, MAGIC can also analyze arbitrarily large samples and is 
robust to changes in the coalescent and recombination processes.
Using MAGIC, we show that the inferred coalescence time histories of 
samples of multiple human genomes exhibit inconsistencies with a description 
in terms of an effective population size based on single-genome data. 
\end{abstract}

\section*{Introduction}

The continuing progress in genetic sequencing technology is enabling
the collection of vast amounts of data on the genomic diversity of populations. 
This data is by far our richest source of information on evolutionary history.
The challenge now is to figure out how to extract this information
-- how to learn as much as possible about the history of populations
from modern data sets of many densely-sequenced individuals.

Perhaps the best-established approach to historical inference from genetic data 
is to fit demographic models to the site frequency spectrum (SFS) (e.g., \cite{gutenkunst2009,excoffier2013}).
The SFS is easy to calculate, even from very large samples,
and demographic models can be fit to it without a specific model of recombination,
but it neglects all information about how diversity is distributed across the genome, 
treating each site independently.
This is a natural approximation for sparse sequencing data, 
where polymorphic sites are generally only very weakly linked, but in 
large samples sequenced at high coverage much of the information
is contained in associations among different polymorphisms.
Because SFS-based approaches cannot use this information, they
can only reliably determine models with a small number of parameters \citep{myers2008,bhaskar2014}.

Recently, an alternative approach has been developed in which
a hidden Markov model is used to explicitly model recombination along the genome 
(the ``sequential Markovian coalescent'', SMC or SMC', \cite{mcvean2005,marjoram2006,paul2011}),
vastly increasing the amount of information that can be gleaned from samples of a small number of individuals
\citep{hobolth2007,li2011,harris2013,sheehan2013,schiffels2014,steinrucken2015}.
But this requires modeling coalescence and recombination throughout the analysis,
and as a result becomes computationally intractable for large samples.
Additionally, for an increasing number of populations, 
we have multiple genomic sequences but know almost nothing about their natural histories, 
including plausible historical demographies and patterns of recombination and selection; 
this is true even for some model organisms
(see \cite{alfred2015} and other articles in series).
It is unclear how accurate these model-based methods are on populations which violate their
underlying assumptions.

Here we present a method for Minimal-Assumption Genomic Inference of Coalescence (MAGIC) 
that infers statistics of the ancestry and history of recombination of an arbitrarily large sample of genomes 
while making only minimal, generic assumptions about recombination, selection, and demography. 
MAGIC finds approximate distributions of times to different common ancestors of the sample. 
These distributions can then be used to fit and test potential models  for the history of the population,
including the simplest model of a  single time-dependent ``effective population size'', $N_e(t)$.
MAGIC strikes a balance between the SFS- and SMC-based approaches, 
using the distribution of diversity across genomic windows of varying size
to generate a description of the single-locus coalescent process
that contains far more information than the simple SFS 
without using a detailed model for recombination.

\section*{Results}

\subsection*{Approach}

\begin{figure}
\includegraphics[width=\linewidth]{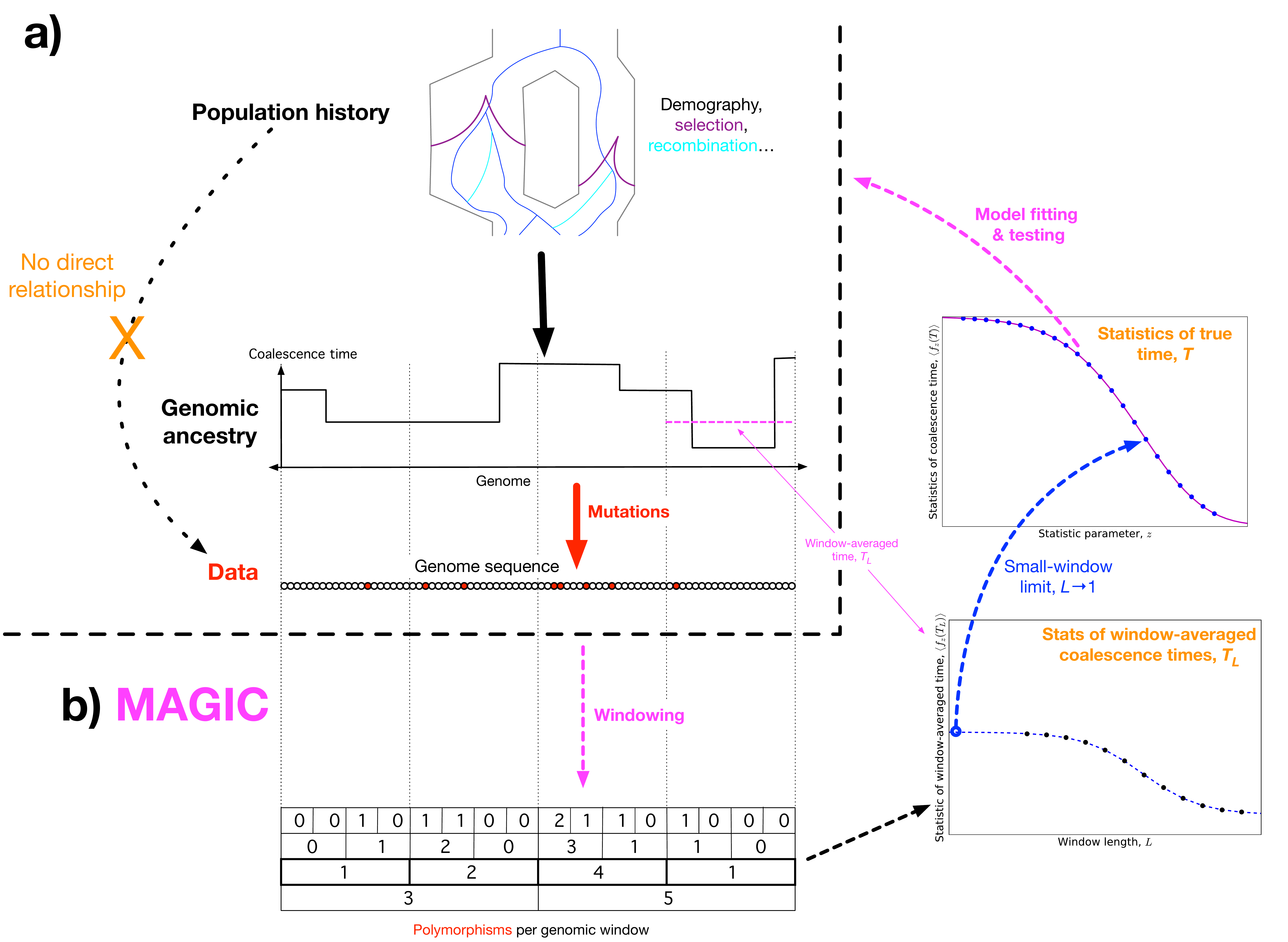}
\caption{\label{fig:schematic_method} 
\textbf{a)} Concept: we would like to infer the history of the population (top)
from the sequence data (bottom), but the causal connection between the two
 is entirely mediated by the 
coalescent history of the sample (middle).
This suggests that it should be possible to extract much of the coalescent 
information from the data without making strong assumptions about the population dynamics.
\textbf{b)} Schematic algorithm: 
MAGIC first splits the sample into small windows and counts the polymorphisms 
within each window, then progressively merges pairs of adjacent windows together 
(bottom left). 
For each window length $\len$, the histogram of window diversity
is used to calculate statistics of the window-averaged coalescence time $\Tl$
(bottom right).
Taking the limit of these as $\len$ goes to 1 gives the statistics of the true 
coalescence time $T$ (top right),
which are then used to fit and test models for the underlying dynamics.
}
\end{figure}

The key fact underlying MAGIC is that the relationship
  between the population parameters (such as recombination rates, historical 
  demography, and selection) and genomic data is entirely mediated by the coalescent history of the sample (\fig{schematic_method}, blue and red boxes).
  We therefore take it as our goal to learn the coalescence time distribution directly from the data without needing a model for the population dynamics.
  Once one knows the coalescent history, the genomic data contains no additional information about the population parameters, 
  and one can fit or evaluate a wide range of models without having to re-analyze the full data set every time \cite{gattepaille2016}.

Essentially, MAGIC uses the variability in the density of polymorphisms across a wide range of length scales
to learn the genome-wide distribution of coalescent histories.
This technique is inspired by \cite{li2011}'s method, PSMC, and its successor MSMC \citep{schiffels2014}, 
which use the fact that SNPs tend to be dense in regions with a long time to the most recent common ancestor (TMRCA),
and sparse in regions with short TMRCAs (\fig{schematic_method}, top left and middle left).
Thus, the distribution of SNPs across the genome can be used to infer the distribution of \textit{local} coalescence times. 
But while PSMC and MSMC use models for coalescence and recombination to assign a coalescence time to
each locus, MAGIC estimates the genomic distribution of times directly,
bypassing the need for explicit modeling.
To do this, MAGIC first splits the genome 
into windows and finds the distribution of genetic diversity across windows, i.e.,  
the histogram of the number of polymorphic sites per window of a given length (\fig{schematic_method}, bottom left; \fig{snpdist}). 
This histogram is then used to infer the statistics of \textit{window-averaged} coalescence times
(\fig{schematic_method}, bottom right; \fig{LTLexample}).
For small windows, these times are essentially the true single-locus coalescence times,
but the inference is noisy due to the small number of mutations in each window. 
For large windows, the inference is more accurate but the window-averaged distribution is far from the 
single-locus distribution because windows typically span multiple linkage blocks. 
The basic trick of MAGIC is that rather than choosing one window length,
it integrates the information gathered from  a wide
 range of different window lengths to find the small-length limit -- the true single-locus distribution 
 (\fig{schematic_method}, center right). 
  
MAGIC's accuracy is comparable to the state-of-the-art model-driven method MSMC \citep{schiffels2014}
on small data sets that conform to MSMC's assumptions (see ``Single diploid samples'' in ``Results''),
but it also enables analyzing data from populations with features (such as gene conversion, ancestral structure, or linked selection)
that violate those assumptions, as well as larger samples.
MAGIC's algorithm, described in ``Methods'' below, is designed to be as simple and modular as possible,
 allowing one to analyze large samples and incorporate additional assumptions
 in situations where more information is available. 
This also enables the inference of a wide range of summary statics, 
including the  distribution of map lengths of blocks of identity-by-descent across the genome
\citep{ralph2013}.
Finally, MAGIC can use the dependence of the window-averaged statistics on window
size to learn about the rate of recombination and the
variation in recombination rate and the coalescent process across the genome. 

\subsubsection*{Representing coalescence time distributions}

For single diploid samples, the coalescent history is completely described by a single time at each locus.
Thus the pair-wise coalescence time distribution could equivalently be described by 
the hazard function (the pairwise coalescence rate, as in \cite{schiffels2014})
or its reciprocal (the ``effective population size'' $N_e(t)$, as in \cite{li2011}).
However, when estimating the distribution from noisy data, 
the procedure that minimizes the error for one description
will not in general minimize the error for the others.
We focus on estimating the distribution,
primarily because it naturally generalizes to arbitrary sets of coalescence tree
branch lengths for larger samples (see below).
This also lets us emphasize that the idea that the coalescent can be described by a single $N_e(t)$ is a model that can be tested.
For plotting, we show cumulative distributions rather than densities  
so that we can plot the actual coalescent histories of samples (black curves in \figs{pairwiseSims}{split}),
which consist of discrete sets of events,
and also because the density estimates are very poorly constrained by the data \citep{ralph2013}.

\subsection*{Single diploid samples}

\begin{figure}
\includegraphics[width=\textwidth]{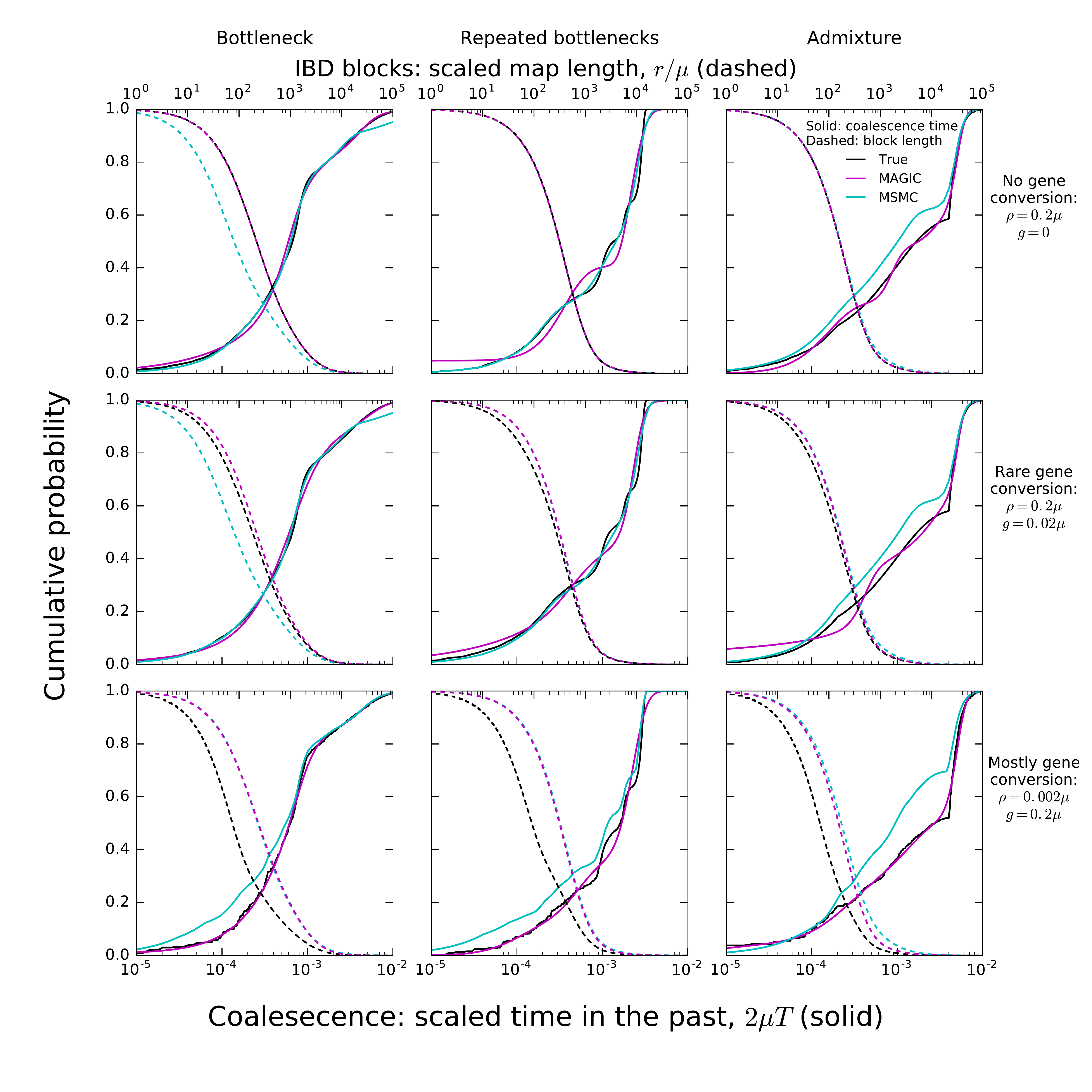}
\caption{\label{fig:pairwiseSims} MAGIC accurately infers the distribution of coalescence times 
(solid curves) and lengths of blocks of identity-by-descent (IBD, dashed curves) for pairwise data simulated with \texttt{ms} under 
several demographic scenarios. 
When crossovers are frequent and gene conversion is rare (top two rows), 
MAGIC and MSMC are comparably accurate for coalescence times.
MAGIC very accurately infers the IBD block length distribution, 
while MSMC sometimes is inaccurate (e.g., ``Bottleneck'' scenario).
For frequent gene conversion and rare crossovers (bottom row),
the details of the gene conversion process have a strong effect 
on the IBD block lengths, and neither method can infer their distribution,
but MAGIC can still infer the coalescence times.
All simulations are of a genome consisting of $100$ independent chromosomes,
each $10^7$ base pairs long, with per-base mutation rate $\mu$ and present population size $N_0$ such that $N_0\mu=10^{-3}$.
Recombination is via crossovers occurring at rate $\rho$ per base, and 
via gene conversion being initiated at rate $g$ per base 
with mean tract length $\lambda = 200\text{bp}$.
See Methods for values of other simulation parameters.
}
\end{figure}

To validate our approach, we have tested MAGIC on single diploid samples generated under a range of coalescent models simulated with \texttt{ms} \citep{hudson2002}.
MAGIC accurately infers the distribution of coalescent times from samples
with map length and polymorphism density similar to that of a human genome (\fig{pairwiseSims}, top two rows, solid curves;
see Methods and \fig{simParams} for detailed parameters).
Although the simulation parameters were conducted under MSMC's assumed model,
MAGIC performs nearly as well (Kolmogorov-Smirnov distance to the true distribution of $4-13\%$ for the simulations shown, 
compared to $4-11\%$ for MSMC).
Both methods tend to smooth out sharp transitions in the coalescence distribution
as a consequence of regularization.
The distribution of map lengths of blocks of identity by descent (IBD)
can be inferred with very high accuracy (\fig{pairwiseSims}, top two rows, dashed curves),
improving on MSMC, which sometimes overestimates the amount of very deep coalescence,
and correspondingly erroneously estimates a large number of very short blocks.
The part of the block-length distribution estimated by MAGIC is complementary to 
the very long blocks that can be observed directly (as in, e.g., \cite{ralph2013}).
MAGIC is also accurate on genomes simulated with \texttt{ms} 
under a model in which recombination is dominated by gene conversion
(\fig{pairwiseSims}, bottom row);
this can be seen as loosely corresponding to a primarily asexual population, 
with gene conversion representing homologous recombination.
In this case, MSMC's recombination model breaks down and MAGIC's 
inferences are more reliable.

\subsection*{Larger samples}

For samples of more than two haplotypes, the coalescent history at each locus
is described by a tree, rather than a single time.
The space of possible trees grows very rapidly with the sample size,
so that even with long genomes it is impossible to directly estimate the full 
distribution.
Instead, MAGIC infers the distribution of some small set of features of the trees, 
such as mean pairwise distance and total branch length,
chosen either because they are important in and of themselves or
because they are sufficient statistics for some model of the coalescent process.
For example, MAGIC can fit the basic time-dependent effective population size model
by estimating the distribution of pairwise coalescence times,
and then check whether the fitted model correctly predicts the distributions of other tree features.

\begin{figure}
\includegraphics[width=\textwidth]{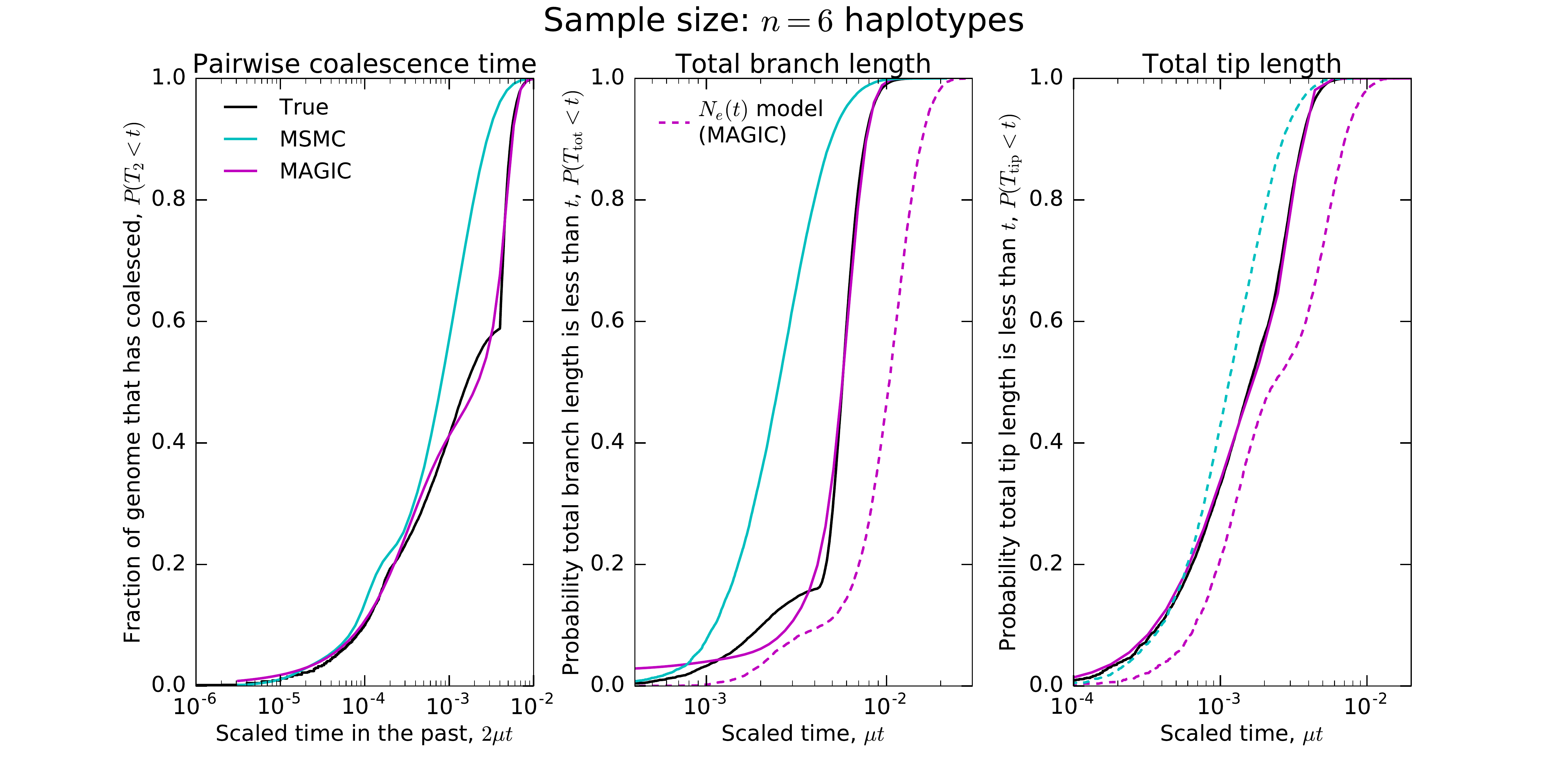}
\includegraphics[width=\textwidth]{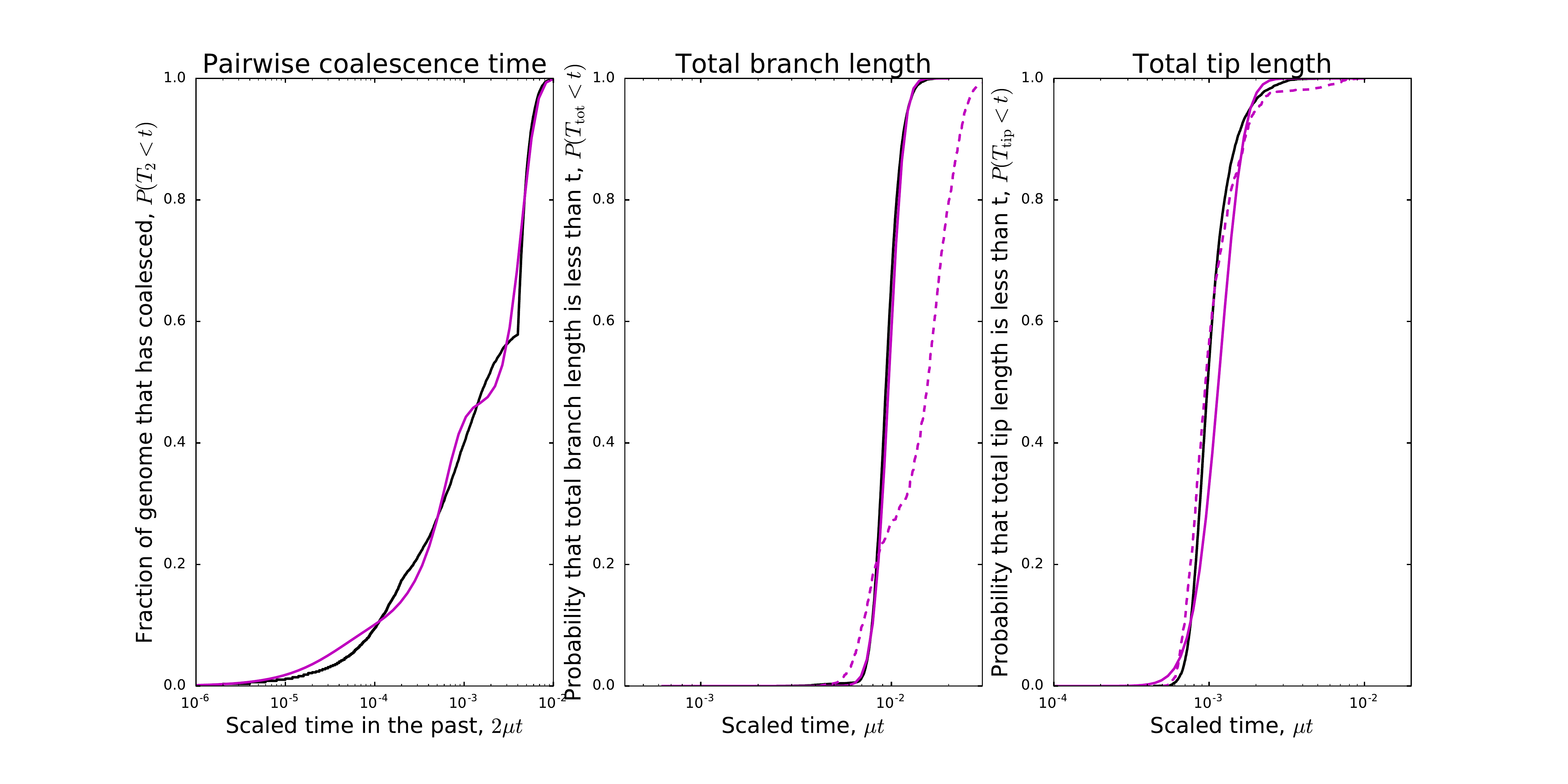}
\caption{\label{fig:split} Coalescence time statistics for a simulated 
population with ancestral structure inferred from larger samples. 
Top row: sample size $n=6$ haplotypes. 
MAGIC accurately estimates the distribution of the pairwise coalescence time,
the total branch length, and the total lengths of the tips of the branches.  
Comparing the estimated distributions reveals that the coalescent process 
cannot be described by an instantaneous rate $1/N_e(t)$.
Bottom row: sample size $n=100$ haplotypes. 
Comparing the pairwise times and the total branch lengths still shows 
strong signs of population structure, but the tip lengths are close
to the pairwise prediction, suggesting that the structure has recently 
disappeared.}
\end{figure}

We test this approach on a sample of 
six haplotypes from a recently admixed population simulated with \texttt{ms}
(demographic parameters in the last column of \figs{pairwiseSims}{simParams}).
MAGIC accurately estimates the distributions of pairwise coalescence times,
the total branch length of the full coalescent trees, 
and the total length of the tips of the coalescent trees (\fig{split}, top row).
We use the first distribution to fit a time-dependent effective population size model,
and then compare its predictions to MAGIC's inferences for the latter two distributions.
The large differences show that the model is
not a good description of the population history.
MSMC's estimate of the pairwise distribution is relatively inaccurate -- perhaps unsurprisingly,
since it is based on an effective population size model.

MAGIC's running time on large samples is dominated by the time to read
all the data through memory, so it grows only linearly with sample size,
meaning that the method can be run on essentially arbitrarily large samples.
MAGIC accurately estimates the distributions of pairwise times,
total branch length, and total tip length in a sample of 100 
haplotypes from the same admixed population (\fig{split}, bottom row).
(Curves are not shown for MSMC because it cannot analyze large samples; 
a sample size of eight from this population caused it to crash.)
The total branch length distribution remains different from that predicted
by the effective population size model, but the tip distribution is close,
differing mainly in the right tail, indicating that the model works better
for recent times than ancient ones, as expected.

In the example above, we have only inferred the distributions of pairwise 
coalescence times, total branch lengths, and total tip lengths.
But to be able to consider a wide range of models for the population,
one must have estimates for a wide range of statistics.
MAGIC can infer the distribution of the total length of any specified set of tree branches.
For a given set, MAGIC first filters the polymorphisms in the original data for 
those that correspond to mutations on the desired branches, 
and then proceeds with the same analysis as in the basic case of a single diploid sample.
For example, to find the distribution of total branch length,
MAGIC analyzes the genomic distribution of all polymorphic sites,
while to find the distribution of total tip length,
it only looks at singletons.
For data with no linkage information, 
the statistics that can be estimated are closely related to the site-frequency spectrum (SFS),
but while the SFS just gives estimates of the \textit{means} of the lengths of different sets of branches,
MAGIC infers full distributions.

\subsection*{Human data}

\begin{figure}
\begin{center}
\includegraphics[width=.7\textwidth]{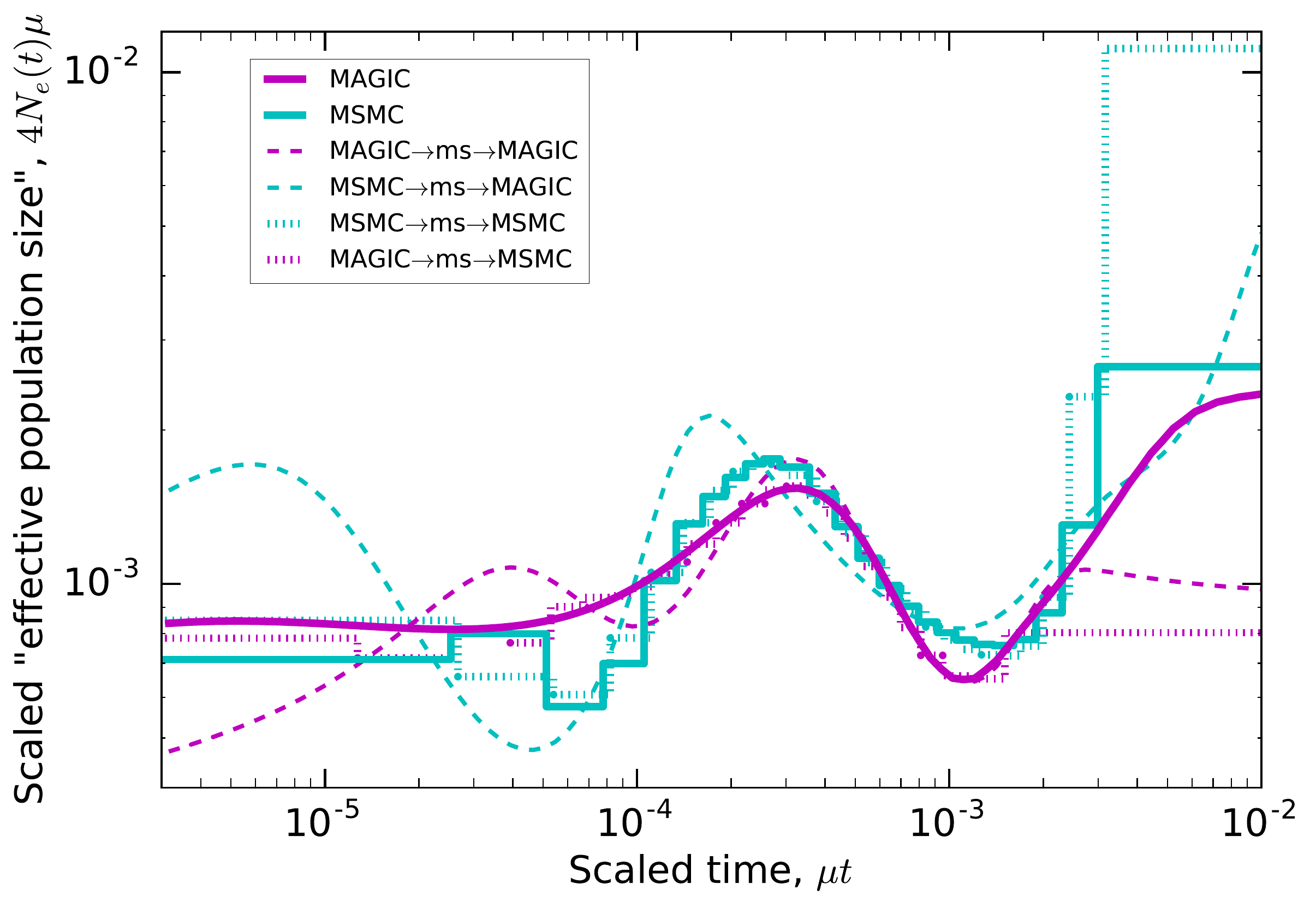}
\includegraphics[width=.7\textwidth]{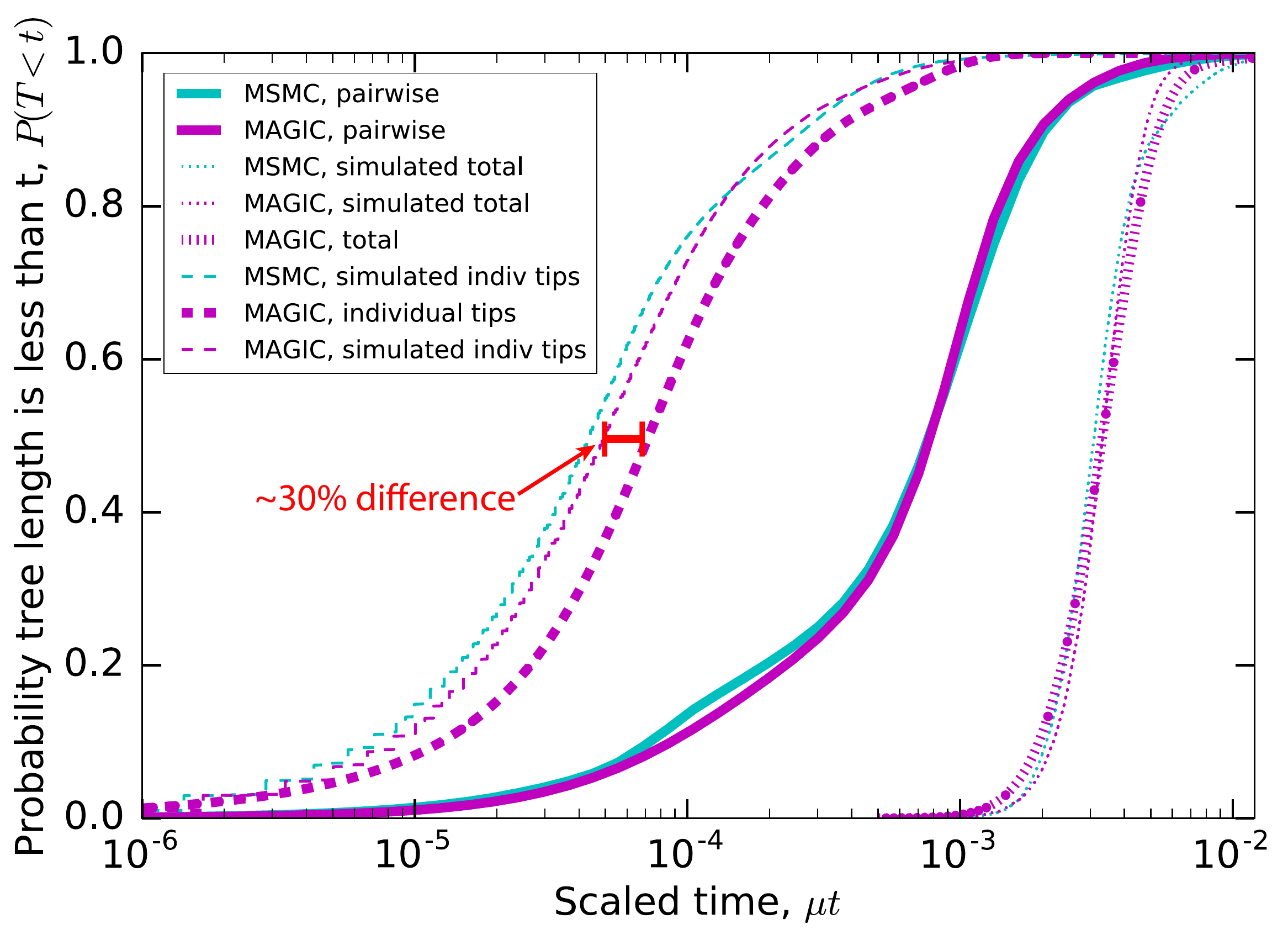}
\caption{\label{fig:NA18502} Inferred evolutionary history of Yoruban individuals.
Top: Inferred ``effective population size'' $N_e(t)$ for YRI individual NA18502. 
MAGIC and MSMC infer similar effective population sizes (solid curves), 
differing mostly in the very recent and distant past where there is limited data.
Running both methods on simulations (dashed and dotted curves) shows that MSMC 
is more accurate in the recent past, while MAGIC may be slightly more accurate in the distant past. 
Bottom: Distributions of coalescence times for a sample of the 9 unrelated Yoruban individuals
from the 69 Genomes data set \citep{drmanac2010}.
These are compared to samples of 9 individuals simulated under the  $N_e(t)$ inferred by MSMC from NA18502 
and by MAGIC from the heterozygosities of all 9 Yorubans (solid curves).  
The pairwise $N_e(t)$ accurately describes the distribution of total branch lengths (dotted curves), 
but underestimates the tip lengths by $\sim 30\%$ (dashed curves).}
\end{center}
\end{figure}

We use MAGIC to analyze human sequences from Complete Genomics' ``69 genomes'' data set \citep{drmanac2010}.
On individual diploid samples, the inferred coalescence time distributions
(plotted in the upper left panel of \fig{NA18502} as ``effective population sizes'')
are similar to those obtained with MSMC, differing mainly in the tails
where the data is limited.
The distribution inferred by MAGIC is closer to that inferred by MSMC using larger
samples (\cite{schiffels2014}, Supplementary Figure 7). 
We also ran MAGIC on a sample comprising all 9 unrelated Yoruban individuals in the data 
set (i.e., a sample of 18 haplotypes, much larger than possible with MSMC)
and compared the inferred tree statistics to those predicted by the Kingman coalescent 
given the inferred pairwise coalescence time distribution (\fig{NA18502}, bottom panel).
The total branch length is close to that predicted by the Kingman model,
but the tips of the trees are substantially longer,
indicating that the inferred effective population size is too small in the recent past,
i.e., it misses the recent population growth.
Thus, by analyzing larger samples, MAGIC is able to probe more recent times than can be 
seen with pairwise comparisons.
However, the observed discrepancy persists back to times at which $>10\%$ of the genome
of single individual has coalesced, by which point there should be more than enough data
to accurately estimate the pairwise coalescence time distribution.
This suggests that the difference is not just due to limited resolution, but 
also to inaccuracies of the $N_e(t)$ model, such that the pairwise coalescent does
not describe the full coalescent process.

\subsection*{Inferring recombination rates}

MAGIC's coalescence-time inference is designed to be robust to 
the form of recombination, but it can also be used to learn about 
recombination.
To do this, rather than simply taking the small-window limit
of the window-averaged coalescence time statistics,
one can look at how they change as a function of window size.
In general, besides the small-window limit in which almost all 
windows lie within IBD blocks, there should also be 
a long-window limit in which almost all windows contain many IBD blocks.
In between these two there is a transitional regime where the window 
length lies within the bulk of the distribution of IBD block lengths;
finding this transitional length gives an estimate of the recombination rate.

Because the transition from the small-window to the large-window limit
is not very sharp, this estimate of the recombination rate is very rough.
A more precise estimation requires a specific model of recombination
and coalescence, like the one used by MSMC. 
But even if one does not have a good model for the dynamics
of a population, one can make the assumption that all autosomes are 
experiencing roughly the same dynamics, whatever they are.
(This assumption is implicit in all demographic inference from full genomes.)
In that case, the dependence of the statistics of the window-averaged coalescence time 
on window length should be similar across autosomes,
and differences in  average recombination rates across chromosomes
should be detectable as rescalings of the window lengths.
MAGIC can therefore use these rescalings to precisely estimate 
\textit{relative} recombination rates.

\begin{figure}
\begin{center}
\includegraphics[width=.45\textwidth]{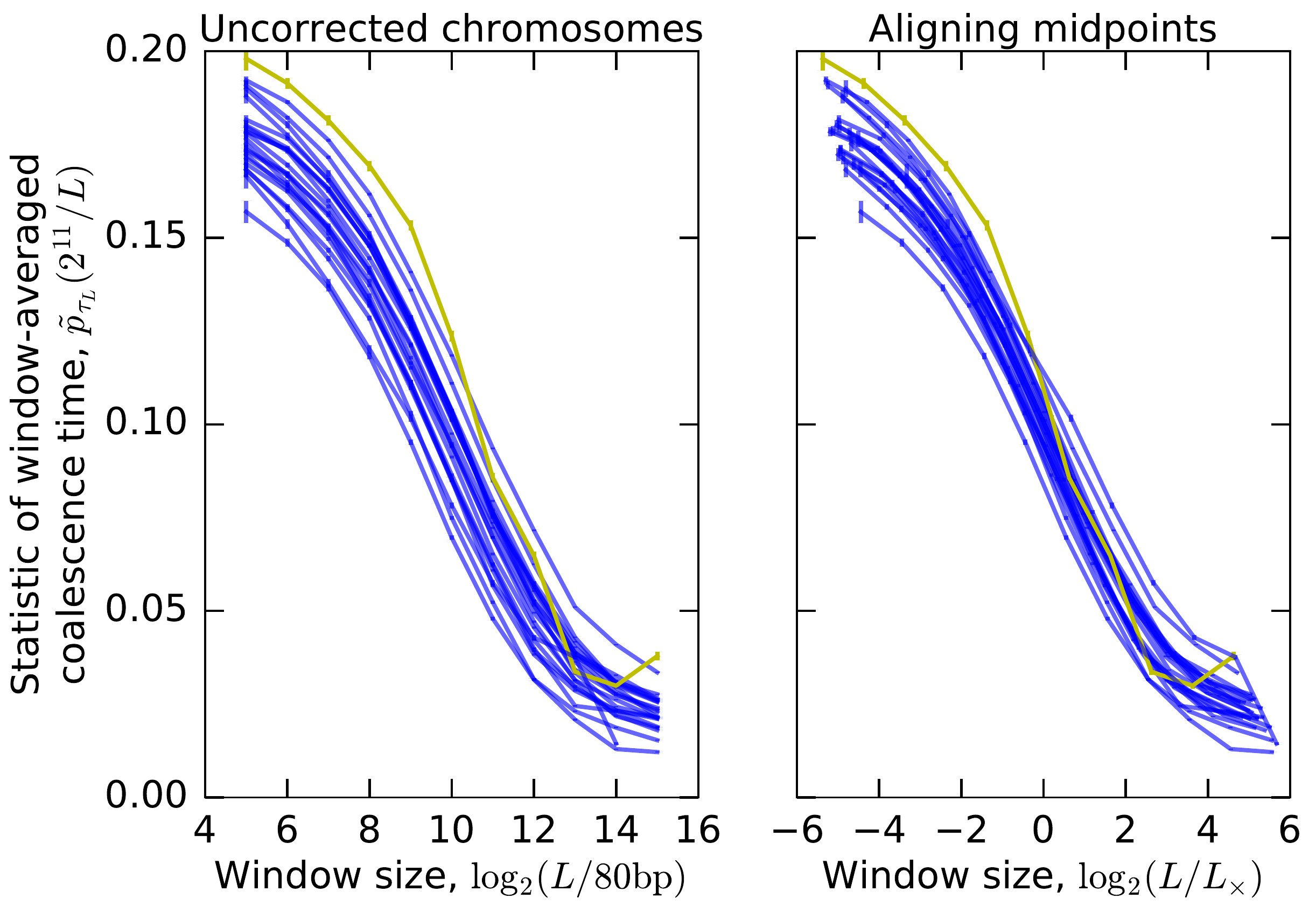}
\includegraphics[width=.45\textwidth]{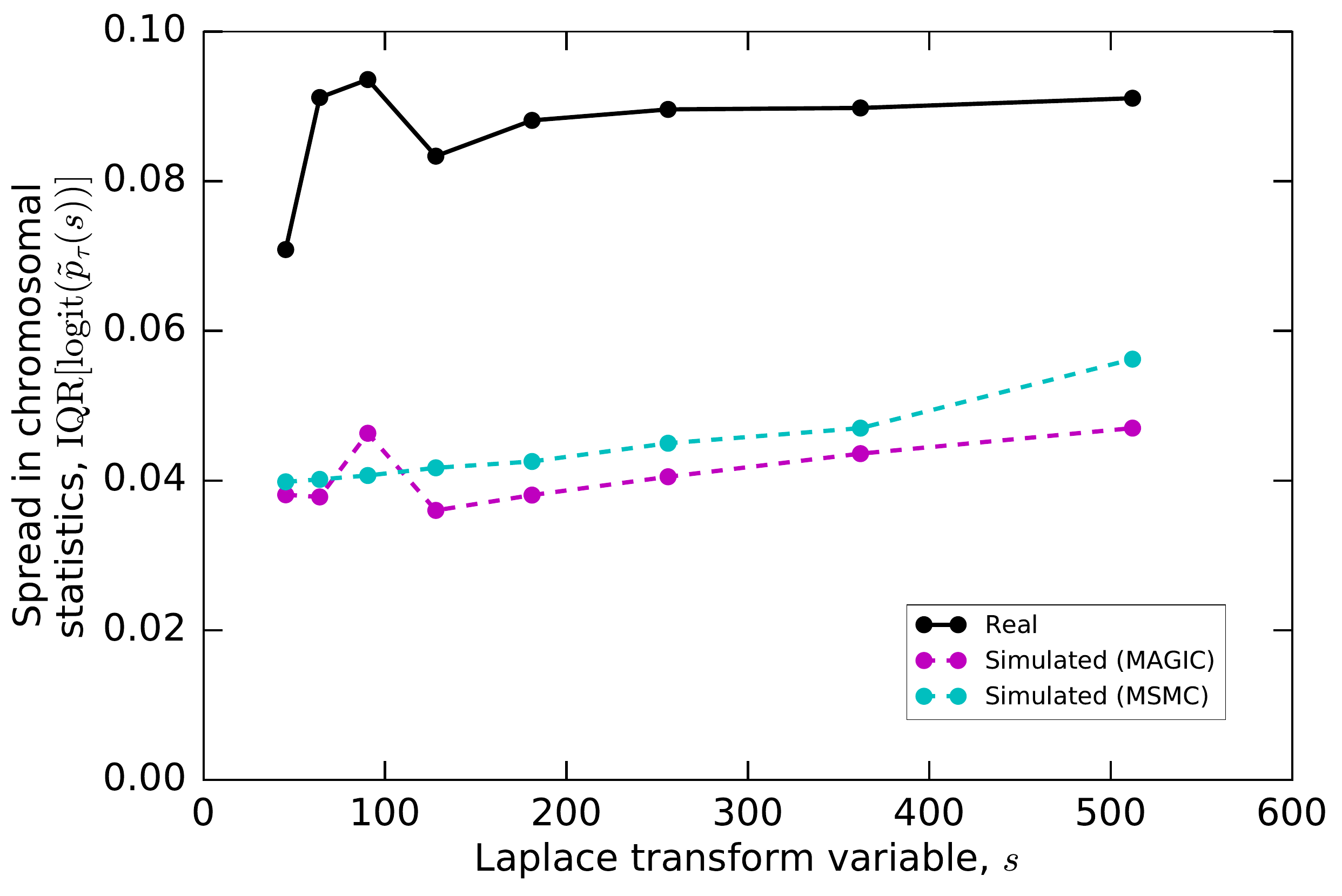}
\includegraphics[width=.8\textwidth]{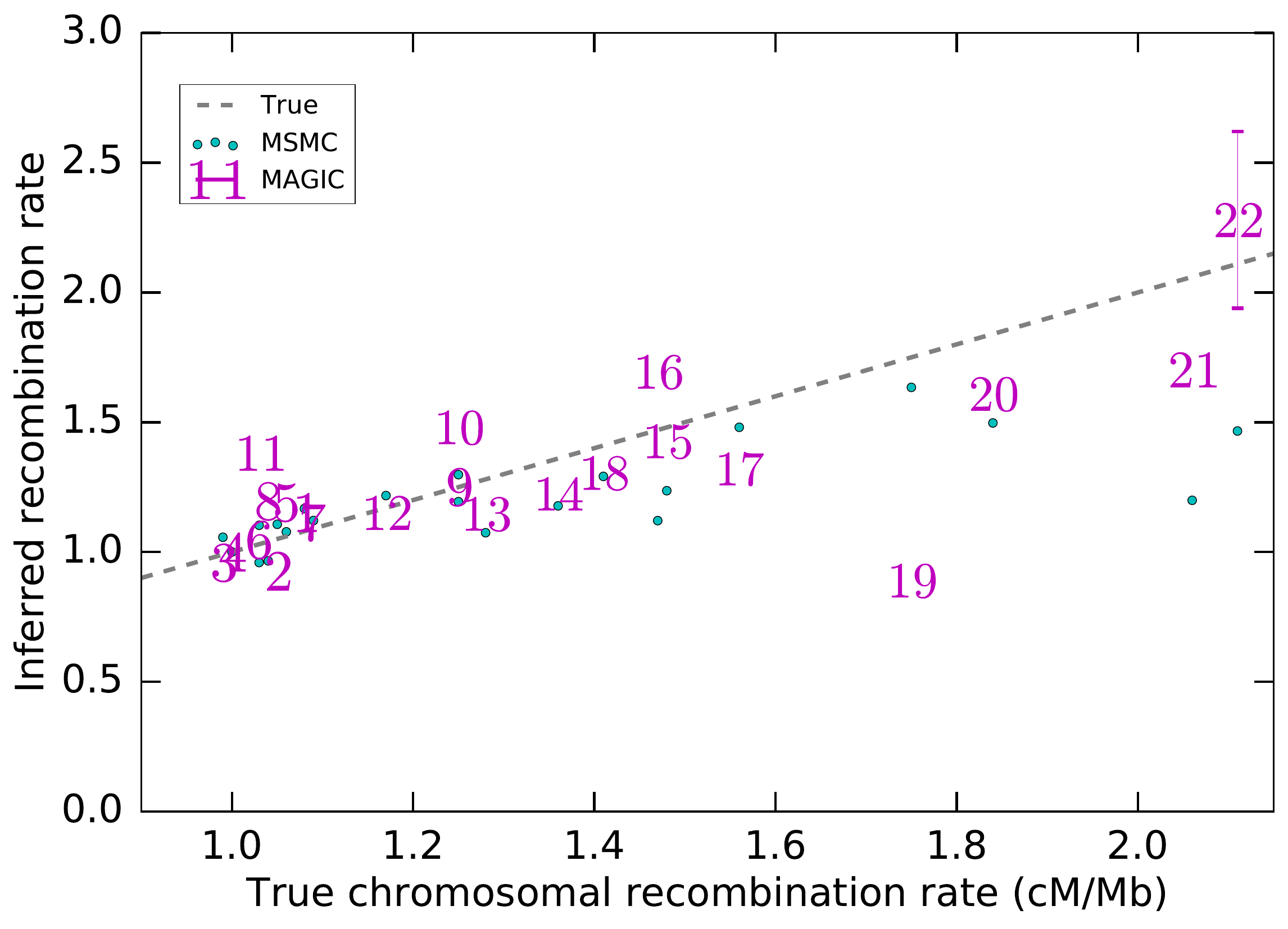}
\end{center}
\caption{\label{fig:chromosomes} 
Estimating relative recombination rates from the genomes of Yoruban individuals.
Top left: One value of the Laplace transform of the window-averaged coalescence time distribution as a function of window size for each autosome.
Top center: Much inter-autosomal variation can be explained by variation in recombination rates:
the curves are similar under a rescaling of window lengths, except
for chromosome 19 (yellow), which appears to have a different pattern of 
coalescence.
Top right: There is still substantial variation in the asymptotes
that cannot be explained by variation in recombination rates,
and is more than expected from intrinsic coalescent stochasticity.
Plot shows the interquartile range of the Laplace transform of the coalescence time distribution across chromosomes for the actual data as well as simulations of the pairwise coalescent histories inferred by MAGIC and MSMC (\fig{NA18502}).
Bottom: The rescaling of window sizes needed to align the different autosomes gives an estimate of their relative recombination rates 
which is very close to the values obtained by \protect\cite{kong2002} (``True'').
For chromosomes other than 22, the inferred error bars are smaller than the 
size of the markers.}
\end{figure}

\begin{figure}
\includegraphics[width=\linewidth]{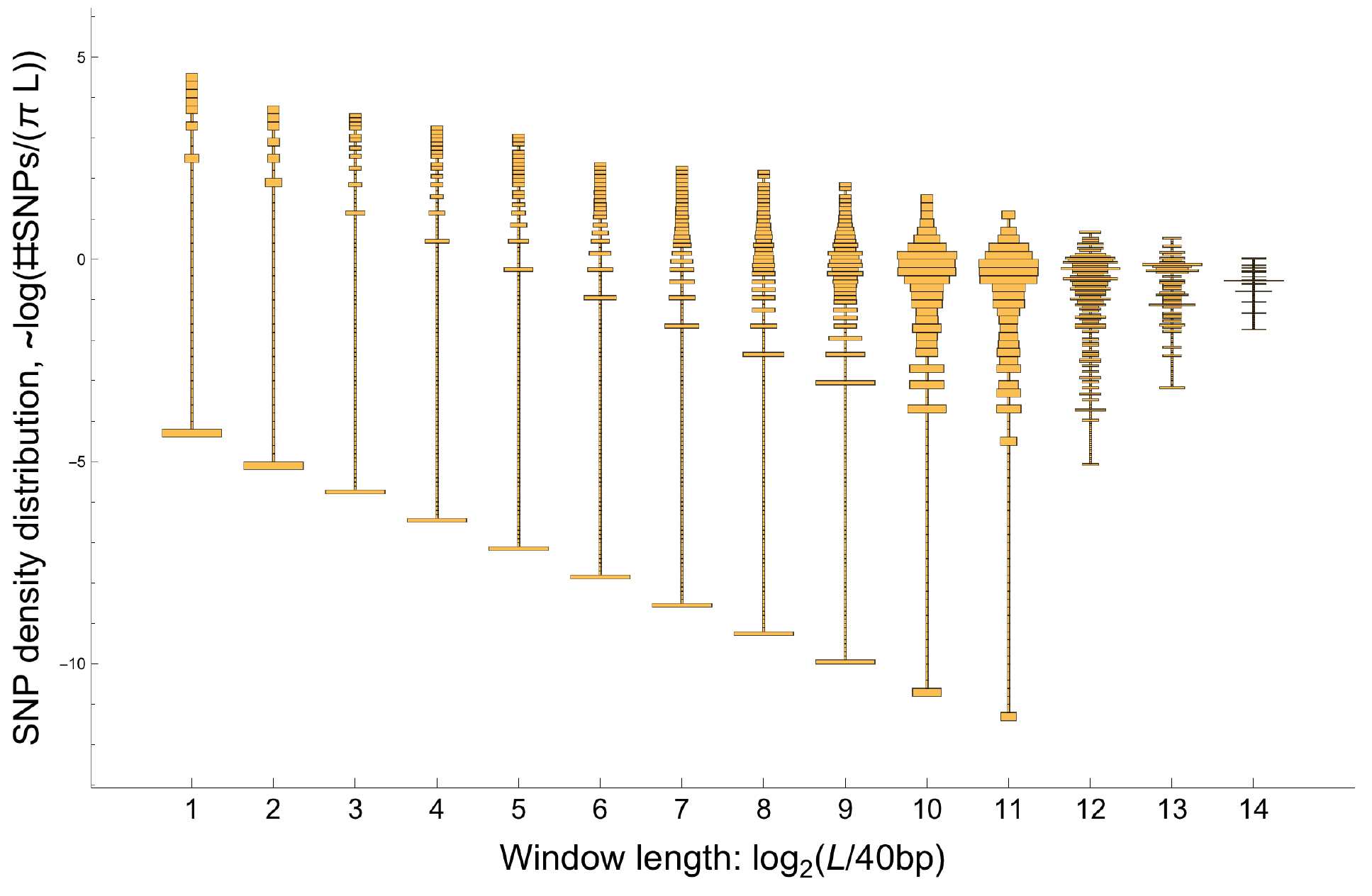}
\caption{\label{fig:snpdist} Distribution of SNP densities across windows at different window lengths, normalized by mean value. 
Data from NA18501 (YRI) chromosome 1.
The bottom bars are windows with no SNPs, which are artificially put at finite values on the log scale.
At short lengths, the distribution is bunched near zero, while at long lengths it is bunched near the mean. 
The best spread is in between, at kb scales.} 
\end{figure}

As an example of this approach, we analyze each autosome 
across the 9 unrelated Yoruban individuals in the data set.
We find that they all show similar dependence of the statistics of the window-averaged coalescence 
time on window length (\fig{chromosomes}, top left).
Up to a rescaling in length, the autosomes appear very similar (\fig{chromosomes}, top center),
with the exception of 19.
The collapse of the remaining 21
autosomes suggests that they differ primarily in  
the amount of very recent coalescence
(rescaling the heterozygosity) and in average recombination rates
(rescaling the window lengths).
The scaling factors for the window lengths therefore are an
estimate of relative recombination rates,
and are indeed very close to values measured by \cite{kong2002} (\fig{chromosomes}, bottom),
with the exception of chromosome 19, as expected.

It is no surprise that chromosome 19 is an outlier in coalesecence: 
it has a much higher gene density than the other chromosomes \citep{grimwood2004},
and is therefore likely to have a much higher fraction of loci under selection and affected by 
 linked selection \citep{hernandez2011}.
However, the other autosomes do not have identical gene densities,
and there are several large regions with unusual patterns of diversity,
such as the MHC locus and flanking regions on chromosome 6.
Indeed, even after rescaling, there is still more residual
variation in coalescence across the 21 similar autosomes
than would be expected by chance (\fig{chromosomes}, top right). 
This variation must be due to non-demographic factors driving 
coalescence; the fact that they are readily detectable suggests that
one should be cautious in interpreting the details of inferences by
MSMC and other demographic inference methods in terms of demography.

\section*{Discussion}

The MAGIC algorithm bridges the gap between fast but limited SFS-based
approaches to demographic inference and 
model-based approaches that are limited to small sample sizes,
allowing far more information to be extracted from large, high-coverage samples.
To see the difference between MAGIC and an SFS-based approach,
consider the information that can be gained from sites with singleton polymorphisms.
Under an approach that treats sites independently, these can be summarized
by one number -- their genomic frequency -- which can only be used to 
estimate one number -- the mean total tip length of coalescent trees.
MAGIC, in contrast, also considers how clustered the sites are over a 
wide range of lengthscales, allowing it to estimate not just the mean,
but the whole distribution of total tip lengths.
MAGIC in this sense is similar in spirit to \cite{bunnefeld2015} 
and \cite{reddy2016}'s ``blockwise SFS''
approach, but differs in that it does not require prior knowledge about
recombination or the range of IBD block lengths.
In addition, because MAGIC converts the genomic distribution of diversity
into the statistics of single-locus coalescent times, it can be checked
or fit with single-locus coalescent simulations, which are much less computationally
intensive than multi-locus ones.

While existing methods rely on fitting simplified demographic models while 
neglecting selection \citep{schraiber2015},
MAGIC estimates makes no assumptions about whether coalescence is driven by 
demography or selection, and only minimal assumptions about mutation and 
recombination. 
We hope that this will make MAGIC useful as a 
first-pass analysis of genomes from species whose natural histories are not 
already well-known, 
with its results informing the choice of more detailed, model-based methods 
that use additional information outside of the sample sequences.
Even for populations for which there are good models, the minimal-assumption approach has advantages.
Because MAGIC has a modular structure and is not tailored to a specific population model, 
it can be used to quickly analyze many populations with very different dynamics,
with each population's model incorporated in just the last step of the analysis.
Similarly, for any given population, MAGIC can calculate many different statistics
describing coalescence and recombination to answer multiple questions about the historical dynamics.
Finally, not using any explicit model of coalescence and recombination
keeps MAGIC's algorithm simple enough that it runs quickly even on very large sample sizes,
and that users familiar with Python can understand and modify it.

There are a number of potential modifications to MAGIC that users could make.
At a minimum, there are likely to be technical improvements to the estimation methods that 
would allow it to get more information out of the data
More interestingly, the range of statistics estimated by MAGIC could be extended.
In particular, MAGIC currently infers the distributions of features of coalescent trees that 
can be found from unphased, unpolarized polymorphism data, but it could be
extended to take advantage of this extra information when available.
It would also be possible to extend MAGIC so that it would infer \textit{joint}
distributions of different coalescence times, rather than just all the marginal
distributions. 
This would greatly increase the amount of information that could be extracted from
extremely large data sets such as are likely to be available in the near future.

\section*{Methods}
\subsection*{Approach}

A sample set of genomes will comprise many blocks of sequence with different coalescent histories;
by looking at the distribution of genetic diversity across blocks,
 one can estimate the coalescence time distribution of the population the sample was drawn from.
\cite{li2011} and \cite{schiffels2014} try to do this by finding the exact boundaries between the blocks using a hidden Markov model.
However, this is only easy to do when mutation rates are much larger than recombination rates, which is generally not the case,
and describing every block becomes impractical for larger sample sizes as the number of blocks proliferates.
Instead, we simply divide the genome into windows of a fixed length $\len$, and consider the distribution of histories of windows.
MAGIC estimates the distribution of a single coalescence time (i.e., coalescence tree statistic) $T$ across genomic positions $x$.
For single diploid samples, $T(x)$ is the total branch length (twice the time to the most recent common ancestor at position $x$) 
and completely characterizes by the coalescent history.
For larger samples, MAGIC can be used to estimate multiple statistics one at a time.

The diversity (e.g., heterozygosity in a single diploid sample) in a window of length $\len$ starting at position $x_0$
depends only on the \textit{window-averaged coalescence time}, $\Tl$ defined as 
\begin{equation}
\Tl \equiv \frac{1}{\len}\sum_{x=x_0}^{x_0+\len-1} T(x).\label{mix}
\end{equation}
If $\len$ is smaller than most block lengths, then windows will typically lie within blocks, 
and the cumulative distribution $\PTl$ of $\Tl$ will be close to the cumulative distribution $\PT$ of $T$. 
For very large $\len$, each window will average over many blocks, and $\PTl$ will have a narrow support around the mean of $\PT$.
Usually, there will be a wide range of intermediate values of $\len$ for which windows lie inside long blocks but cover multiple short blocks.

Given that $\PTl$ approaches $\PT$, as $\len$ decreases, one might be tempted to take $\len$ to be as small as possible,
 but the problem of course is that we cannot see $\Tl$ directly; we need to infer it from the number of SNPs in the window.
Given $\Tl$, and assuming a constant mutation rate $\mu$ per base, the number of SNPs will be approximately Poisson-distributed with mean $\mu\len\Tl$.
(Here we are assuming that $\mu T$ is always small enough that each base has only a small chance of having mutated;
this allows us to approximate the underlying sum of binomially-distributed random variables with a Poisson that depends only on $\Tl$.)
The smaller $\len$ is, the lower the signal-to-noise ratio will be and the less power we will have to distinguish different values of $\Tl$. 
Thus, we expect that we will get the most information about $\PT$ from an intermediate value of $\len$, 
and should be able to do better still by integrating information from multiple values of $\len$.

The total probability that there are $n$ SNPs in a window of length $\len$ is by averaging the Poisson distribution over all possible values of $\Tl$:
\begin{equation}
\Pl(n) = \evgo{e^{-\taul} \taul^n}{\taul}/n!,\label{Pln}
\end{equation}
where $\taul$ is the window-averaged coalescence time scaled such that it is equal to the
expected number of SNPs in the window: $\taul\equiv\mu\len\Tl$.
$\Pl$ is a Poisson mixture distribution, with mixing distribution given by $\Ptaul$,
the cumulative distribution function of $\taul$, i.e., the fraction of the genome that is expected to have coalesced by a given scaled time.
Our immediate goal is to estimate $\PTl$ from the observed SNP count distribution $\est{\Pl}$.

If only a small, random fraction of the genome has been sequenced,
this is the classic statistical question of estimating the mixing 
distribution of a Poisson mixture from a finite number of draws; 
in this case, the maximum likelihood estimate for $\Ptaul$ is a step function 
with a number of steps 
equal to roughly half the number of distinct values of $n$ (number of SNPs in a window)
observed across the genome \citep{simar1976}.
This maximum likelihood function is difficult to find, however,
and in any case we are typically in 
a slightly different situation, in which most of the genome has been sequenced 
and the sampling noise in the number of windows with a given $\Tl$ is small.
More importantly, we are interested primarily in the true distribution $\PT$
rather than the window-averaged distribution $\PTl$,
so we only want to estimate features of $\PTl$ that also describe $\PT$.

\subsection*{Laplace transforms}

Rather than trying to estimate the full distribution $\PTl$,
we will instead estimate a set of statistics describing it.
The Laplace transform $\ltptaul(z) \equiv \evg{e^{-z\taul}}$,
evaluated at a set of points $\{z_j\}$, is a natural choice,
as it is closely related to the diversity distribution:
\eqref{Pln} shows that $(-1)^n \Pl(n)$ is the $n^\text{th}$
Taylor coefficient of $\ltptaul(z)$ about $z=1$.
This has two implications.
First, the similarity to the proportion of homozygous windows
$\Pl(0)=\evg{e^{-\taul}}$ means that the Laplace transform
has a natural interpretation as an estimate for the proportion
of windows of length $\len$ that would be homozygous if the mutation rate were 
multiplied by $z$. (It is also closely related to the distribution of lengths of IBD blocks -- see below.)
Second, we can quickly and easily estimate $\ltptaul$
using the plug-in estimator:
\begin{equation}
\est{\ltptaul(z)} = \sum_{n=0}^\infty \est{\Pl(n)}(1-z)^n.\label{lt}
\end{equation}

The estimate \eqref{lt} of the Laplace transform will be accurate for $z$ close to 1, but will 
blow up for large $z$ -- we cannot accurately estimate the amount of very recent coalescence.
To make this precise, we need to estimate the error in \eqref{lt}.
We use \cite{ghosh1983}'s formula (5) to estimate the value of $\Tl$ 
for every window from the observed number of SNPs, 
with the adjustment that for the $K_0$ windows with 0 SNPs, 
we use $\est{\Tl}=\log(2)/(\mu K_0)$ rather than 0.
We then estimate the error introduced by the stochastic 
mutation accumulation process under the resulting $\est{\PTl}$.
The accuracy of both \eqref{lt} and the error estimate could be improved by using more sophisticated estimators, 
but the current ones are the easiest to compute 
and appear to give reasonable values for all the simulated data that we have tried.

\subsection*{Combining lengthscales}

\begin{figure}
\includegraphics[width=\linewidth]{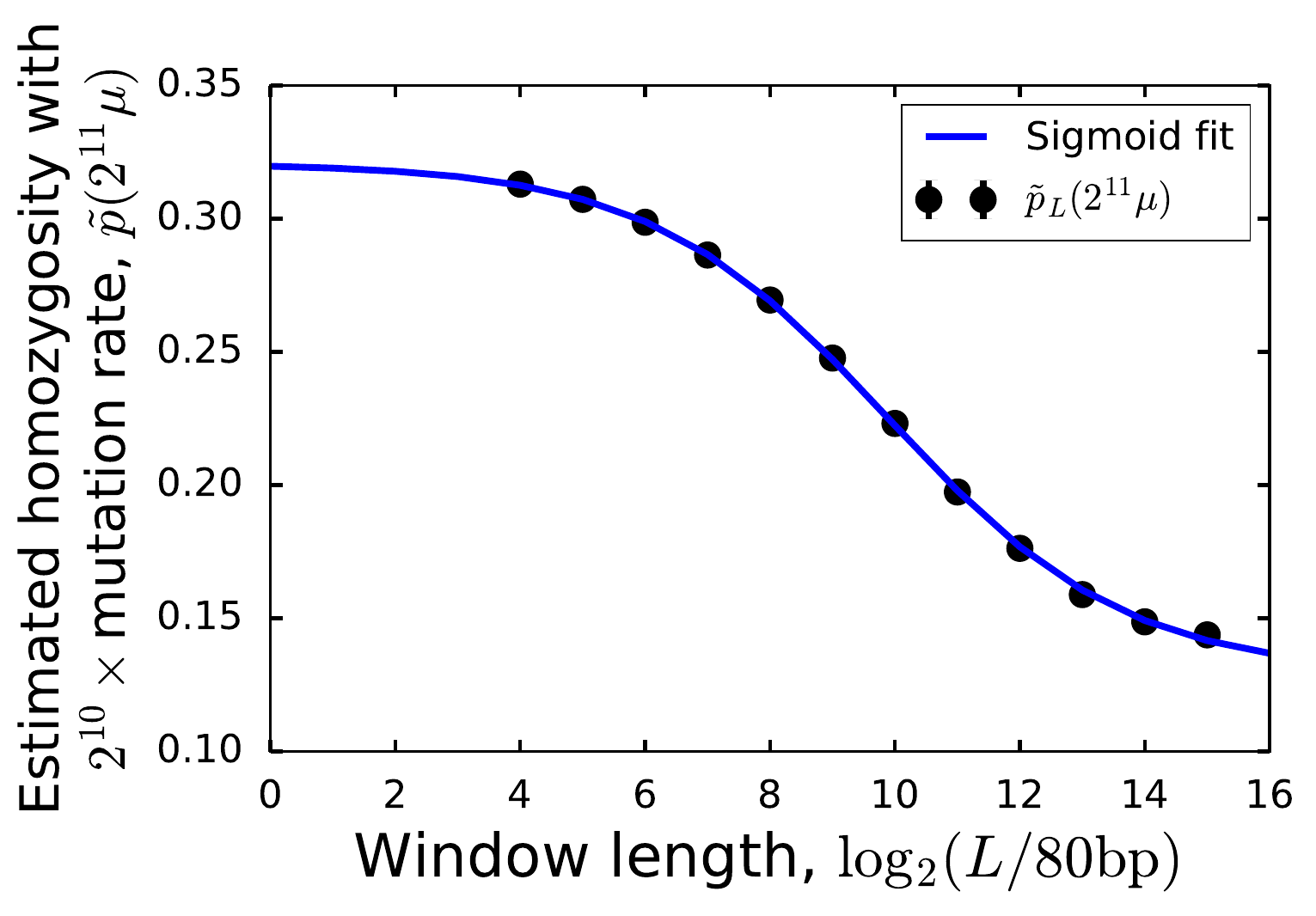}
\caption{\label{fig:LTLexample} 
The window-averaged coalescence time distribution changes with window size. 
For short windows, the distribution approaches the true coalescence time distribution,
while for long windows it approaches a distribution sharply peaked at the mean coalescence time. 
The length $\sim 5\text{kb}$ at which it begins to deviate from the small-window limit gives an estimate of the density of recombination events.
Data is from YRI individual NA18502, chromosome 1.}
\end{figure}

To combine information from different window lengths, we need to correct for the 
increase in window-wide mutation rate $\mu \len$. 
We can therefore consider the quantity $\ltptaul(z/\len)$ as a function of $\len$,
holding $z$ fixed, as shown in \fig{LTLexample}.
When $\PTl$ is nearly independent of $\len$, this quantity should be nearly constant.
(To see this, note that $\ltptaul(z/\len) = \evg{e^{-z\mu \Tl}}$, with no explicit $\len$ dependence.)
This is the case for very large $\len$, when each window averages over many coalescent blocks,
and for very small $\len$, where each window falls within a coalescent block
and $\PTl$ approaches $\PT$, the distribution we are interested in.
We therefore fit a sigmoid curve (specifically, Richards' curve) 
to $\est{\ltptaul}(z/\len)$ as a function of $\log(\len)$,
\begin{equation}
\ltptaul(z/\len) \approx a_z + \frac{b_z - a_z}{\left[1 + \left(\len/\len_\times\right)^{-c_z}\right]^{1/\nu_z}}, \text{ with } a_z, b_z \in [0,1];\, c_z, d_z, \nu_z > 0;\, \len_\times > 1, \label{sigmoid}
\end{equation}
and take the left asymptote $a_z$ as an estimate of $\ltpT(\mu z)$.
The right asymptote $b_z$ is the long-window limit, and can therefore be estimated 
directly from the genomic density of SNPs $\varpi$: $b_z=e^{-\varpi z}$.
To fit the remaining parameters, we us \texttt{curve\_fit} function in SciPy's \texttt{optimize} package.
\texttt{curve\_fit} also returns a standard error $\est{\sigma^2(\ltpT(\mu z))}$,
estimated under the assumption that the errors in $\est{\ltptaul(z/\len)}$ are independent 
for different values of $\len$. 
This is obviously not true (all estimates are from the same set of mutations),
but even for lengthscales that are only separated by factors of $2$,
the correlations in the error appear to be small in simulations,
giving final error bars of roughly the right magnitude.
While $a_z$ contains the information about the single-locus coalescent,
the values of other parameters, particularly $\len_\times$, are informative about
recombination and the relationships among loci -- see ``Inferring recombination rates''
in main text.

The sigmoid form \eqref{sigmoid} is flexible enough to fit all of the simulated and real 
data that we have examined, but
we only trust our estimate of $\ltpT(\mu z)$ if the data are
 close enough to the left asymptote  so that the 
estimate is not very sensitive  to the choice of functional form. 
Effectively, this means that the estimate is close to $\est{\ltptaul(z/\len)}$
for the smallest $\len$ for which the estimated error bars are small, 
with corrections based on the next few higher length scales. 
But this smallest $\len$ depends on $z$, so while for any given value of $z$
only a few lengthscales are important, 
every lengthscale is important for estimating the Laplace transform for some $z$.
Short lengthscales are useful for small $z$ (i.e., long coalescence times),
while long lengthscales are useful for large $z$ (short coalescence times).

\subsection*{Coalescence-time distributions}

Once we have estimates for the Laplace transform of the coalescence time distribution
at a set of $\{z_j\}_{j=1,\ldots,J}$,
we would like to invert the transform to obtain $\pT$.
Unfortunately, this is a fundamentally hard problem \citep{epstein2008},
and we need to use some kind of regularization.
We do this by assuming that $\PT$ can be written as a mixture
of gamma distributions:
\begin{equation}
\pT(t) = \sum_{i=1}^{\lfloor(J+1)/3\rfloor} c_i \frac{t^{k_i-1}e^{-t/\theta_i}}{\Gamma(k_i)\theta_i^{k_i}},
\end{equation}
where $\sum_i c_i =1$ and all $c_i$, $k_i$, and $\theta_i$ are positive.
This can also be extended to include a possible point mass at $t=0$
when estimating the distribution of features that we expect to be exactly 0
for some coalescent trees, e.g., the length of branches that are ancestral to 
5 haplotypes out of a sample of 10.

\begin{figure}
\includegraphics[width=\linewidth]{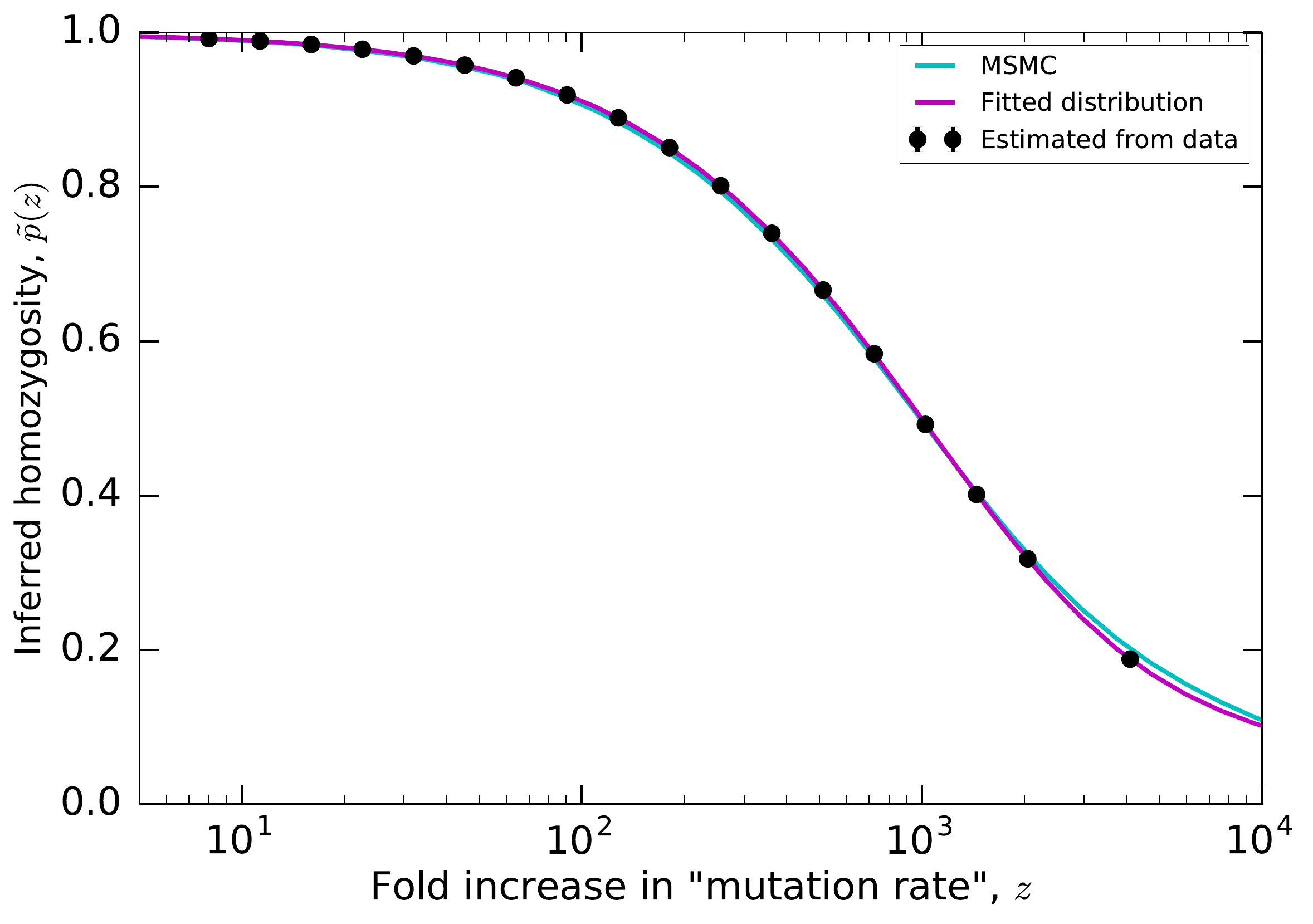}
\caption{\label{fig:LTexample} Laplace transform of the coalescence time distribution. 
Points are values estimated using the scale-dependence of the window-averaged diversity distributions,
as shown in \fig{LTLexample}. 
Magenta curve shows MAGIC's fitted gamma mixture distribution, 
cyan curve shows the Laplace transform of MSMC's estimated coalescence time distribution. 
The two curves are close, but differ slightly for very large $z$, corresponding to very recent times.
}
\end{figure}

The gamma mixture distribution has the computational advantage of having a simple analytic
Laplace transform:
\begin{equation}
\ltpT(z) = \sum_{i=1}^{\lfloor(J+1)/3\rfloor} c_i (1+\theta_i z)^{k_i},\label{gammaLT}
\end{equation}
with an additional constant term if a point mass at $t=0$ is included.
This means that we can simply fit \eqref{gammaLT} to the values
$\left\{\est{\ltpT(z_j)}\right\}$ without having to deal
with inverse Laplace transforms directly.
We use the L-BFGS-B algorithm implemented in SciPy
to find the $\{c_i,k_i,\theta_i\}$ that minimize
the scaled squared error
$\sum_j \left(\ltpT(z_j)-\est{\ltpT(z_j)}\right)^2/\est{\sigma^2(\ltpT(z_j))}$.
The gamma mixture form is flexible enough to fit all the data that we have tried (see, e.g., \fig{LTexample}).

\subsubsection*{Block-length distributions}

Identical-by-descent (IBD) blocks are stretches of the genome that have not undergone recombination since the common ancestor of the block.
The distribution of block lengths can be used to infer patterns of relatedness and ancestry \citep{li2011,ralph2013},
but it is hard to measure except for long blocks with very recent common ancestors,
because the recombination events are not directly observable.
Under the standard assumptions that coalescence is mostly driven by neutral processes (rather than linked selection) and that recombination primarily occurs via crossovers,
the distribution of the genetic map lengths of these blocks across the genome is closely related to the Laplace transform of the coalescence time distribution: 
\begin{equation}
P(\rblock>r) = -\ltpT'(r)/\ev{T} = \ltpT'(r)/ \ltpT'(0).\label{blocklength}
\end{equation}
We can estimate the block-length distribution in Morgans simply by approximating the derivative of $\ltpT$,
a much easier problem than inverting the transform. 
If we want to convert map lengths to numbers of bases, we need to estimate the crossover frequency per base.
The dependence of $\ltptaul$ on $\len$ (i.e., $\len_\times$ in \eqref{sigmoid} above) provides a rough estimate; 
MSMC gives a more precise one.
\eqref{blocklength} gives the distribution of lengths of blocks that are not interrupted by even ``ghost'' recombination events
where the coalescent tree does not change (\cite{marjoram2006}'s ``\textit{R} class'' of events).
In the simplest case of a single diploid sample from a well-mixed population evolving neutrally under the Kingman coalescent,
these ghost events can be ignored by simply scaling all recombination rates by $2/3$, 
but in general they must be included, even though they are not directly observable.

\subsection*{Implementation}

The code for MAGIC is written in Python and is available at https://github.com/weissmanlab/magic-beta.
It uses the same input format as MSMC and the \texttt{msmc-tools} suite.

\subsection*{Data processing}

We use the 69 Genomes Diversity Panel from Complete Genomics \citep{drmanac2010}, 
and use msmc-tools \citep{schiffels2014} to turn the data into a list of SNPs. 
We split the genome into windows of 80bp, count the number of SNPs in each window,
and then repeatedly merge all windows in pairs and re-count to get the SNP
count distribution at successively larger length scales (\figs{schematic_method}{snpdist}).
(To correct for uneven sequencing coverage across windows, all windows with $<80\%$ coverage
were dropped, and all with $>80\%$ coverage were down-sampled to $80\%$.)
This gives us SNP count distributions at a range of length scales for every 
chromosome of every individual in the data set.

\subsection*{Simulated data}

All coalescent simulations were done in \texttt{ms} \citep{hudson2002}.
To make the simulations computationally tractable, 
genomes were assembled from independently simulated ``chromosomes'' 
of $10^7$ bases each.

\begin{figure}
\begin{center}
\includegraphics[height=4in]{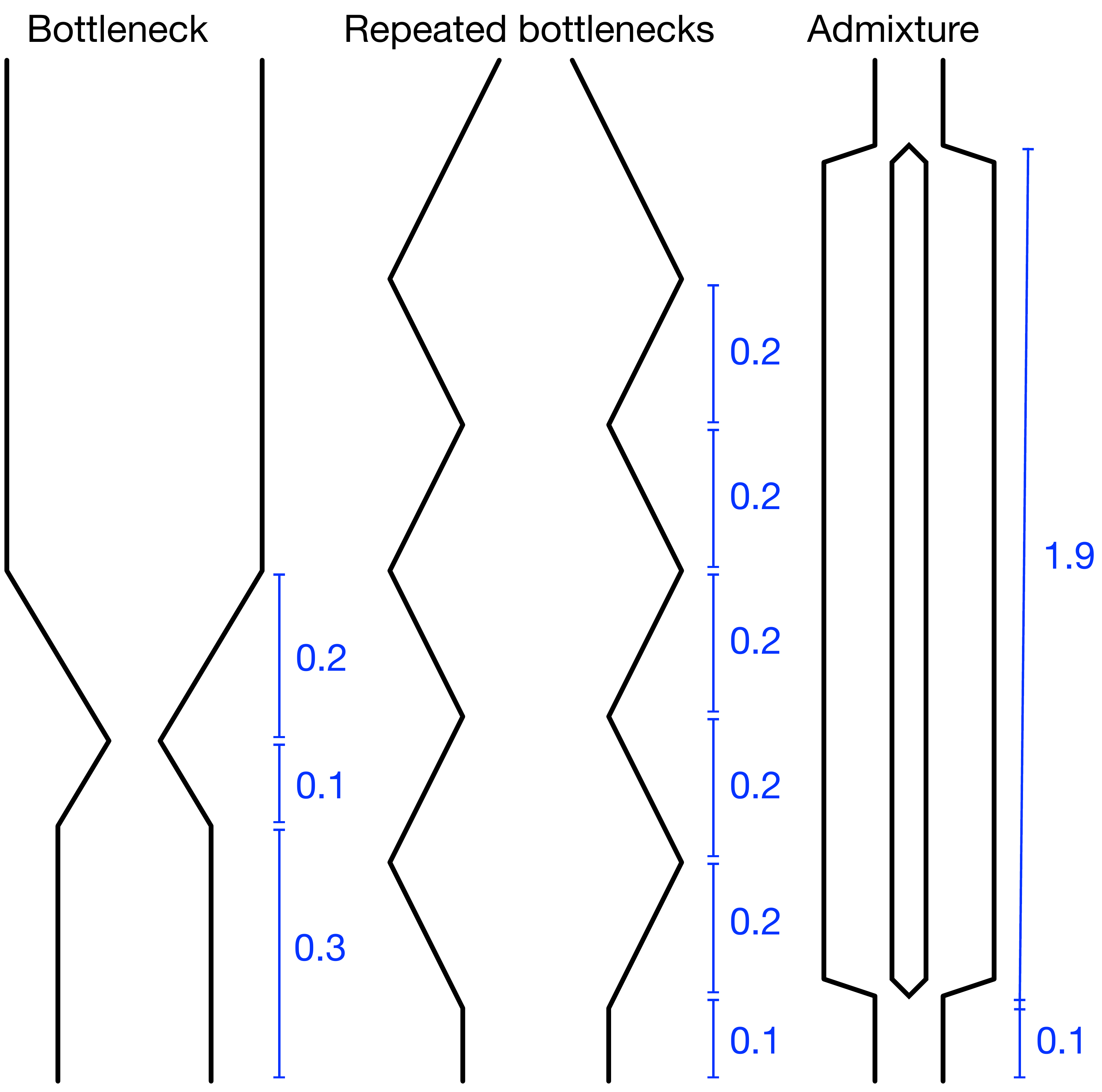}
\caption{\label{fig:simParams} Demographic scenarios simulated. 
All time intervals are in units of $4N_0$.
In the ``bottleneck'' and ``repeated bottlenecks'' scenarios,
the population grows and shrinks exponentially at rate $10/(4N_0)$.}
\end{center}
\end{figure}

For the test demographic scenarios in \figs{pairwiseSims}{split},
the per-base mutation rate was $\mu = 10^{-3}/(4N_0)$.
(\texttt{ms} is parametrized in terms of the present 
population size, $2N_0$.)
For the ``bottleneck'' scenario, the demography was given by the command ``\texttt{-eG .3 10 -eG .4 -10 -eG .6 0}''; 
for ``repeated bottlenecks'', by ``\texttt{-eG .1 -10 -eG .3 10 -eG .5 -10 -eG .7 10 -eG .9 -10 -eG 1.1 10 }''; 
and for ``admixture'', by ``\texttt{-es .1 1 .5 -ej 2 1 2}''.
See \fig{simParams} for schematics.
For the pairwise simulations, each sample consisted of $100$ chromosomes 
with recombination rates as listed in 
\fig{pairwiseSims}. 
For the larger-sample simulations in \fig{split}, 
each sample consisted of $10$ chromosomes
with per-base rate of crossovers $\rho=\mu/5$,
and per-base rate of initiation of gene conversion $g=\mu/20$
with mean tract length $\lambda = 1\text{kb}$.

\section*{Acknowledgements}

We thank Stephan Schiffels for help with MSMC, 
Kelley Harris and Rasmus Nielsen for help with human population genetic data,
Peter Ralph for discussions of the mathematical analysis,
and Razib Khan for suggesting the name of the method.
This work was supported by the National Institute Of General Medical Sciences of the National Institutes of Health under Award Number R01GM115851 (O.H.) and by a Simons Investigator award from the Simons Foundation (O.H.).

\clearpage

\bibliography{magic}

\begin{thebibliography}{10}
\providecommand{\url}[1]{\texttt{#1}}
\providecommand{\urlprefix}{URL }
\expandafter\ifx\csname urlstyle\endcsname\relax
  \providecommand{\doi}[1]{doi:\discretionary{}{}{}#1}\else
  \providecommand{\doi}{doi:\discretionary{}{}{}\begingroup
  \urlstyle{rm}\Url}\fi
\providecommand{\bibAnnoteFile}[1]{%
  \IfFileExists{#1}{\begin{quotation}\noindent\textsc{Key:} #1\\
  \textsc{Annotation:}\ \input{#1}\end{quotation}}{}}
\providecommand{\bibAnnote}[2]{%
  \begin{quotation}\noindent\textsc{Key:} #1\\
  \textsc{Annotation:}\ #2\end{quotation}}
\providecommand{\eprint}[2][]{\url{#2}}

\bibitem{gutenkunst2009}
Gutenkunst RN, Hernandez RD, Williamson SH, Bustamante CD (2009) Inferring the
  joint demographic history of multiple populations from multidimensional {SNP}
  frequency data.
\newblock PLoS Genetics 5: e1000695.
\bibAnnoteFile{gutenkunst2009}

\bibitem{excoffier2013}
Excoffier L, Dupanloup I, Huerta-Sanchez E, Sousa VC, Foll M (2013) Robust
  demographic inference from genomic and {SNP} data.
\newblock PLoS Genetics 9: e1003905.
\bibAnnoteFile{excoffier2013}

\bibitem{myers2008}
Myers S, Fefferman C, Patterson N (2008) Can one learn history from the allelic
  spectrum?
\newblock Theoretical Population Biology 73: 342--348.
\bibAnnoteFile{myers2008}

\bibitem{bhaskar2014}
Bhaskar A, Song YS (2014) Descartes' rule of signs and the identifiability of
  population demographic models from genomic variation data.
\newblock The Annals of Statistics 42: 2469--2493.
\bibAnnoteFile{bhaskar2014}

\bibitem{mcvean2005}
McVean GAT, Cardin NJ (2005) Approximating the coalescent with recombination.
\newblock Philosophical Transactions of the Royal Society B: Biological
  Sciences 360: 1387--1393.
\bibAnnoteFile{mcvean2005}

\bibitem{marjoram2006}
Marjoram P, Wall JD (2006) Fast ``coalescent'' simulation.
\newblock BMC Genetics 7: 16.
\bibAnnoteFile{marjoram2006}

\bibitem{paul2011}
Paul JS, Steinr{\"u}cken M, Song YS (2011) An accurate sequentially {M}arkov
  conditional sampling distribution for the coalescent with recombination.
\newblock Genetics 187: 1115--1128.
\bibAnnoteFile{paul2011}

\bibitem{hobolth2007}
Hobolth A, Christensen OF, Mailund T, Schierup MH (2007) Genomic relationships
  and speciation times of human, chimpanzee, and gorilla inferred from a
  coalescent hidden {M}arkov model.
\newblock PLOS Genetics 3: e7.
\bibAnnoteFile{hobolth2007}

\bibitem{li2011}
Li H, Durbin R (2011) Inference of human population history from individual
  whole-genome sequences.
\newblock Nature 475: 493--496.
\bibAnnoteFile{li2011}

\bibitem{harris2013}
Harris K, Nielsen R (2013) Inferring demographic history from a spectrum of
  shared haplotype lengths.
\newblock PLoS Genetics 9: e1003521.
\bibAnnoteFile{harris2013}

\bibitem{sheehan2013}
Sheehan S, Harris K, Song YS (2013) Estimating variable effective population
  sizes from multiple genomes: a sequentially {M}arkov conditional sampling
  distribution approach.
\newblock Genetics 194: 647--662.
\bibAnnoteFile{sheehan2013}

\bibitem{schiffels2014}
Schiffels S, Durbin R (2014) Inferring human population size and separation
  history from multiple genome sequences.
\newblock Nature Genetics 46: 919--925.
\bibAnnoteFile{schiffels2014}

\bibitem{steinrucken2015}
Steinr{\"u}cken M, Kamm JA, Song YS (2015) Inference of complex population
  histories using whole-genome sequences from multiple populations.
\newblock bioRxiv : 026591.
\bibAnnoteFile{steinrucken2015}

\bibitem{alfred2015}
Alfred J, Baldwin IT (2015) New opportunities at the wild frontier.
\newblock eLife 4: e06956.
\bibAnnoteFile{alfred2015}

\bibitem{gattepaille2016}
Gattepaille L, G{\"u}nther T, Jakobsson M (2016) Inferring past effective
  population size from distributions of coalescent times.
\newblock Genetics : in print.
\bibAnnoteFile{gattepaille2016}

\bibitem{ralph2013}
Ralph PL, Coop G (2013) The geography of recent genetic ancestry across
  {Europe}.
\newblock PLoS Biology 11: e1001555.
\bibAnnoteFile{ralph2013}

\bibitem{hudson2002}
Hudson RR (2002) Generating samples under a {W}right-{F}isher neutral model of
  genetic variation.
\newblock Bioinformatics 18: 337--338.
\bibAnnoteFile{hudson2002}

\bibitem{drmanac2010}
Drmanac R, Sparks AB, Callow MJ, Halpern AL, Burns NL, et~al. (2010) Human
  genome sequencing using unchained base reads on self-assembling {DNA}
  nanoarrays.
\newblock Science 327: 78--81.
\bibAnnoteFile{drmanac2010}

\bibitem{kong2002}
Kong A, Gudbjartsson DF, Sainz J, Jonsdottir GM, Gudjonsson SA, et~al. (2002) A
  high-resolution recombination map of the human genome.
\newblock Nature Genetics 31: 241--247.
\bibAnnoteFile{kong2002}

\bibitem{grimwood2004}
Grimwood J, Gordon LA, Olsen A, Terry A, Schmutz J, et~al. (2004) The {DNA}
  sequence and biology of human chromosome 19.
\newblock Nature 428: 529--535.
\bibAnnoteFile{grimwood2004}

\bibitem{hernandez2011}
Hernandez RD, Kelley JL, Elyashiv E, Melton SC, Auton A, et~al. (2011) Classic
  selective sweeps were rare in recent human evolution.
\newblock Science 331: 920--924.
\bibAnnoteFile{hernandez2011}

\bibitem{bunnefeld2015}
Bunnefeld L, Frantz LAF, Lohse K (2015) Inferring bottlenecks from genome-wide
  samples of short sequence blocks.
\newblock Genetics 201: 1157--1169.
\bibAnnoteFile{bunnefeld2015}

\bibitem{reddy2016}
Reddy CB, Hickerson MJ, Frantz LAF, Lohse K (2016) Approximate likelihood
  inference of complex population histories and recombination from multiple
  genomes.
\newblock bioRxiv : 077958.
\bibAnnoteFile{reddy2016}

\bibitem{schraiber2015}
Schraiber JG, Akey JM (2015) Methods and models for unravelling human
  evolutionary history.
\newblock Nature Reviews Genetics .
\bibAnnoteFile{schraiber2015}

\bibitem{simar1976}
Simar L (1976) Maximum likelihood estimation of a compound {Poisson} process.
\newblock The Annals of Statistics 4: 1200--1209.
\bibAnnoteFile{simar1976}

\bibitem{ghosh1983}
Ghosh M, Hwang JT, Tsui KW (1983) Construction of improved estimators in
  multiparameter estimation for discrete exponential families.
\newblock The Annals of Statistics 11: 351--367.
\bibAnnoteFile{ghosh1983}

\bibitem{epstein2008}
Epstein CL, Schotland J (2008) The bad truth about {L}aplace's transform.
\newblock SIAM Review 50: 504--520.
\bibAnnoteFile{epstein2008}

\end{thebibliography}

\end{document}